\documentclass[12pt,preprint]{aastex}

\newcommand{\ts}{\thinspace}
\newcommand{\simless}{\mathbin{\lower 3pt\hbox
     {$\rlap{\raise 5pt\hbox{$\char'074$}}\mathchar"7218$}}}
\newcommand{\simgreat}{\mathbin{\lower 3pt\hbox
     {$\rlap{\raise 5pt\hbox{$\char'076$}}\mathchar"7218$}}}

\newcommand{\about}    {$\sim$\ts}
\newcommand{\aboutless}{$\simless$\ts}

\newcommand{\msun}{\ts M$_\odot$}
\newcommand{\etal}{et~al.}
\newcommand{\IRAS}{{\it IRAS}}
\newcommand{\Hipparcos}{{\it Hipparcos}}
\newcommand{\Spitzer}{{\it Spitzer}}

\begin{document}

\title{Evidence for Mass-dependent Circumstellar Disk Evolution\\
in the 5 Myr-old Upper Scorpius OB Association}

\author{John M. Carpenter\altaffilmark{1}}
\author{Eric E. Mamajek\altaffilmark{2}}
\author{Lynne A. Hillenbrand\altaffilmark{1}}
\author{Michael R. Meyer\altaffilmark{3}}

\altaffiltext{1}{California Institute of Technology, Department of Astronomy, MC 105-24, Pasadena, CA 91125}
\altaffiltext{2}{Harvard-Smithsonian Center for Astrophysics, 60 Garden St.,
MS-42, Cambridge, MA 02138}
\altaffiltext{3}{Steward Observatory, The University of Arizona,
933 North Cherry Ave., Tucson, AZ 85721}

\begin{abstract}

We present 4.5, 8, and 16\micron\ photometry from the {\it Spitzer Space 
Telescope} for 204 stars in the Upper Scorpius OB Association. The data are 
used to investigate the frequency and properties of circumstellar disks around 
stars with masses between \about 0.1 and 20\msun\ at an age of \about 5~Myr. 
We identify 35 stars that have emission at 8\micron\ or 16\micron\ in excess 
of the stellar photosphere. The lower mass stars (\about 0.1-1.2\msun) appear 
surrounded by primordial optically thick disks based on the excess emission 
characteristics. Stars more massive than \about 1.8\msun\ have lower 
fractional excess luminosities suggesting that the inner \about 10~AU of the
disk has been largely depleted of primordial material. None of the G and F 
stars (\about 1.2-1.8\msun) in our sample have an infrared excess at 
wavelengths $\le$16\micron. These results indicate that the mechanisms for 
dispersing 
primordial optically thick disks operate less efficiently on average for low 
mass stars, and that longer time scales are available for the buildup of 
planetary systems in the terrestrial zone for stars with masses 
\aboutless 1\msun.

\end{abstract}


\section{Introduction}

Most young (\about 1\,Myr) stars embedded within molecular clouds are
surrounded by circumstellar accretion disks \citep{Strom89} that are potential
sites of planet formation. The ubiquity of disks extends to all masses from 
as high as 10\msun\ down through brown dwarfs, and in all environments from 
isolated stars in Taurus to dense clusters in Orion \citep{Lada00,Bouy06}.

By an age of 10\,Myr, the primordial disks so ubiquitous around 
young stars change dramatically. The inner disk (\aboutless 1\,AU) dissipates 
in $>$90\% of stars \citep{Mamajek04}, accretion rates drop by an 
order of magnitude \citep{Muzerolle00}, and the mass contained in 
small dust grains declines by at least a factor of four 
\citep{Liu04,Carpenter05}. Results from the {\it Spitzer Space Telescope}
\citep{Werner04} demonstrate an even more 
striking degree of evolution, as dust within a \about 1\,AU orbital radius is 
found in only a few percent of \about 10\,Myr stars 
\citep{Silverstone06}. MIPS\,24\micron\ surveys have detected dust in 
7--48\% of 10--30\,Myr stars \citep{Young04,Rieke05,Chen05}, but with 
fractional dust luminosities orders of magnitude below that found in younger 
sources. Together these observations have established that circumstellar disks 
are at an advanced evolutionary stage by an age of \about 10\,Myr.

Key to understanding the formation of planetary systems is examining the 
evolution of circumstellar disks after the main accretion phase has 
terminated. To measure the properties of disks during this epoch for stellar 
masses ranging from 0.1 to 20\msun, we have conducted a photometric survey of 
205 stars with spectral types between M5 and B0 in the 5 Myr-old Upper 
Scorpius OB association using the IRAC, IRS, and MIPS 
instruments on \Spitzer. This Letter presents analysis of the IRAC and 
IRS photometry to probe for terrestrial-zone material across the stellar mass 
spectrum at a constant age.

\section{Sample Selection}
\label{sample}

The parent sample for this program was selected from previous membership
studies of the Upper Sco OB association. We compiled members identified based 
on (a) \Hipparcos\, astrometry \citep[B, A, F, and G stars;][]{deZeeuw99}, 
(b) optical color--magnitude diagrams and spectroscopic verification of 
lithium \citep[G, K, and M stars;][]{Preibisch99,Preibisch02}, and (c) x--ray 
sources subsequently verified as lithium-rich, pre-main-sequence stars 
\citep[G, K, \& M stars;][]{Walter94,Martin98,Preibisch98,Kunkel99,Kohler00}.
Since these studies identified Upper Sco members based on stellar properties
(proper motion, strong lithium, x--ray emission) rather then those linked to 
circumstellar material (e.g. H$\alpha$ emission, near-infrared excess), we 
believe that our sample is not biased for or against the presence of a 
circumstellar disk.

The parent sample was cross--matched with the \Hipparcos\, \citep{Perryman97}, 
Tycho--2 \citep{Hog00}, and UCAC2 \citep{Zacharias04} proper motion catalogs
where possible. Using the \citet{Madsen02} kinematic model for Upper~Sco, we 
computed the probability that a given star has a proper motion consistent with
membership in the association \citep[see e.g.][]{Mamajek02}.  Any star that 
deviated more than 2$\sigma$ from the proper motion model was removed, as was 
any star with an inferred cluster parallax distance more than 45\,pc from the 
mean Upper Sco distance \citep[where the line-of-sight depth of the 
association is $\sim$30\,pc; ][]{Preibisch99}. We also removed stars located 
in projection against the $\rho$~Oph molecular cloud, which is 
near Upper~Sco and contains stars with ages of \aboutless 1\,Myr. These 
criteria yielded 341 Upper Sco members with spectral type M5 and earlier. 

The aim was to populate five quasi-logarithmically-spaced mass bins with 
50 stars each. In paring the list, we (a) removed stars 
requiring $>20$~cycles with MIPS to detect the photosphere at 24\micron, 
(b) dropped sources where a nearby star compromised the 2MASS photometry,
(c) removed stars with the highest 70\micron\ background levels, and
(d) avoided sources observed by other \Spitzer\ programs.
The final source list consists of 205 stars\footnote{One source, HIP~80112, was
observed only with MIPS and is not further discussed.}: 
48 stars with masses between 0.1 and 0.2\msun\ (corresponding to spectral 
types of \about M3-M5), 50 between 0.2 and 0.4\msun\ (M0.5-M3), 
42 between 0.4 and 1.8\msun\ (F0-M0.5), 
50 between 1.8 and 3.0\msun\ (B5-F0), and 15 more massive 
than 3\msun\ (earlier than B5). 
The final source list does not constitute
a complete sample of stars, but should be a representative population 
of Upper Sco.

\citet{Preibisch99} estimated an age of 5 Myr for a x-ray-selected sample
of stars in Upper Sco as inferred from \citet{DM94} pre-main sequence 
evolutionary tracks after allowing for 
binaries. This age is consistent with the nuclear \citep[5-6~Myr;][]{deGeus89} 
and dynamical \citep[4.5~Myr;][]{Blaauw91} age of the high mass stars. 
Moreover, \citet{Preibisch99} find that the intrinsic age spread within the 
association is less than 2 Myr. We therefore adopt an age of 5~Myr for Upper 
Sco, but recognize that the age is uncertain by at least 1-2 Myr depending on 
the choice of model evolutionary tracks.

\section{Observations and Data Reduction}
\label{obs}

IRAC \citep[4.5\micron\ and 8\micron,][]{Fazio04} observations for 204 stars 
and IRS peak-up-imaging \citep[PUI; 16\micron,][]{Houck04} data for 195 stars 
were obtained with the {\it Spitzer Space Telescope}. IRS PUI observations 
were not attempted for nine B-stars since the detector would have saturated on 
the stellar photosphere. Exposure times ranged from 0.02 to 12 seconds for 
IRAC and from 6 to 30 seconds for IRS PUI depending on the stellar brightness 
estimated from 2MASS photometry. At least nine dither positions were obtained 
per band, and the number was increased as needed to achieve a minimum signal 
to noise ratio on the stellar photosphere of 50 for IRAC and 20 for IRS PUI. 

Data analysis was performed on the Basic Calibrated Data images 
produced by the S14 pipeline for IRAC and S13 for IRS PUI. Photometry 
was measured on individual frames using a modified version of IDLPHOT. For 
IRAC, we adopted an aperture radius of 3 pixels (1 pixel = 1.22\arcsec) and 
a sky annulus between 10 and 20 pixels. For IRS PUI, we used an aperture 
radius of 2 pixels (1 pixel = 1.8\arcsec) and a sky annulus between 5 and 8 
pixels. A multiplicative aperture correction of 1.110, 1.200, and 1.316 was 
applied to the IRAC 4.5\micron, IRAC 8\micron, and IRS 16\micron\ flux densities
to place the photometry on the calibration scale described in the IRAC and IRS 
data handbooks. Photometric corrections at the few percent level were 
applied to the IRAC data to account for distortion and variations in the 
effective bandpass across the detector \citep{Reach05}.
Internal photometric uncertainties were computed as standard deviation of the 
mean of measurements made on individual frames.  We adopted a minimum 
uncertainty of 1.22, 0.66, and 0.58\% for 4.5\micron, 8\micron, and 
16\micron\ respectively based on repeatability achieved for bright stars.

We incorporate into the analysis 14 solar-type stars in Upper Sco from the FEPS
\Spitzer\ Legacy Program \citep{Meyer06} that were selected for that study 
based on criteria similar to those stated in \S\ref{sample}. We exclude 
the FEPS source HD~143006 since this star was recognized as a Upper Sco
member based on an \IRAS\ excess \citep{Odenwald86} and thus would bias the 
sample. The FEPS IRAC data were processed using the above procedures.
FEPS did not obtain IRS PUI observations. 

Table~\ref{tbl:phot} lists the sources, spectral types, \Spitzer\ 
fluxes, and internal uncertainties for the 218 stars analyzed here. The 
uncertainties do not include calibration uncertainties of 2\% for 
IRAC \citep{Reach05} and 6\% for IRS PUI as quoted in the IRS data handbook
\footnote{The calibration factors adopted here are 0.1388 and 0.2021
MJy/sr per DN/s for IRAC 4.5\micron\ and 8\micron\ and 
0.01375 MJy/sr per $e^{-1}$/sec for IRS 16\micron\ PUI.}. Five sources are
flagged in Table~\ref{tbl:phot} where the curve of growth at 16\micron\ 
deviates from a point source by more than 4\% for the adopted aperture radius, 
indicating that the flux measurement may include contributions from a second
source.

\section{Sources with Infrared Excesses}

Color-color diagrams for the Upper Sco sources are presented in 
Figure~\ref{fig:ccd}. The top panel shows the 8\micron\ to 4.5\micron\ flux 
ratio ($\equiv R_8$) as a function of the $J-H$ color, and the bottom panel 
shows the 16\micron\ to 4.5\micron\ flux ratio ($\equiv R_{16}$). In both 
panels, most sources lie along a tight locus that is assumed to represent 
emission dominated by reddened stellar photospheres. However, several
sources have large values of $R_8$ or $R_{16}$ diagnostic of 8\micron\ or 
16\micron\ emission in excess of the photosphere.

Since the scatter in the observed colors is likely dominated by factors not 
easily quantified on a star-by-star basis, we determined empirically a 
threshold to identify sources with intrinsic infrared excesses. A linear 
relation was fitted between log~$R_{16}$ and $J-H$. Any outliers more distant 
than four times the RMS of the fit residuals were removed, and the fit was 
repeated until no additional outliers were identified. A similar fit was 
performed between log~$R_8$ and $J-H$ after removing all $R_{16}$ outliers.

The RMS residuals from the final linear fit were 1.9\% and 5.7\% for 
$R_8$ and $R_{16}$ respectively. A source was identified with an infrared 
excess if $R_8$ or $R_{16}$ exceeded the fitted relation by both four 
times the RMS of the fit residuals and four times the internal uncertainty 
in the flux ratio. We further required that a large value of $R_8$ or 
$R_{16}$ does not result from extinction as determined from $B$, 
$V$ and 2MASS photometry, spectral types, and the \citet{Mathis90} 
extinction law. 

Excesses were inferred toward 35 sources as indicated in Table~\ref{tbl:phot}: 
29 at 8\micron\ and 33 at 16\micron. 
The 16\micron\ excess sources include $[$PBB2002$]$\,USco\,J161420.2$-$190648, 
which saturated the IRS detector. HIP\,78207 and 
$[$PZ99$]$\,J161411.0$-$230536 have 8\micron\ excesses but were not observed 
at 16\micron. An IRS spectrum of the latter source shows a clear excess at 
this wavelength \citep{Carpenter07}, and HIP\,78207 exhibits \IRAS\ excesses 
at 12\micron\ and 25\micron\ \citep{Oudmaijer92}. 

\section{Discussion}

\begin{figure}
\includegraphics[angle=0,scale=0.7]{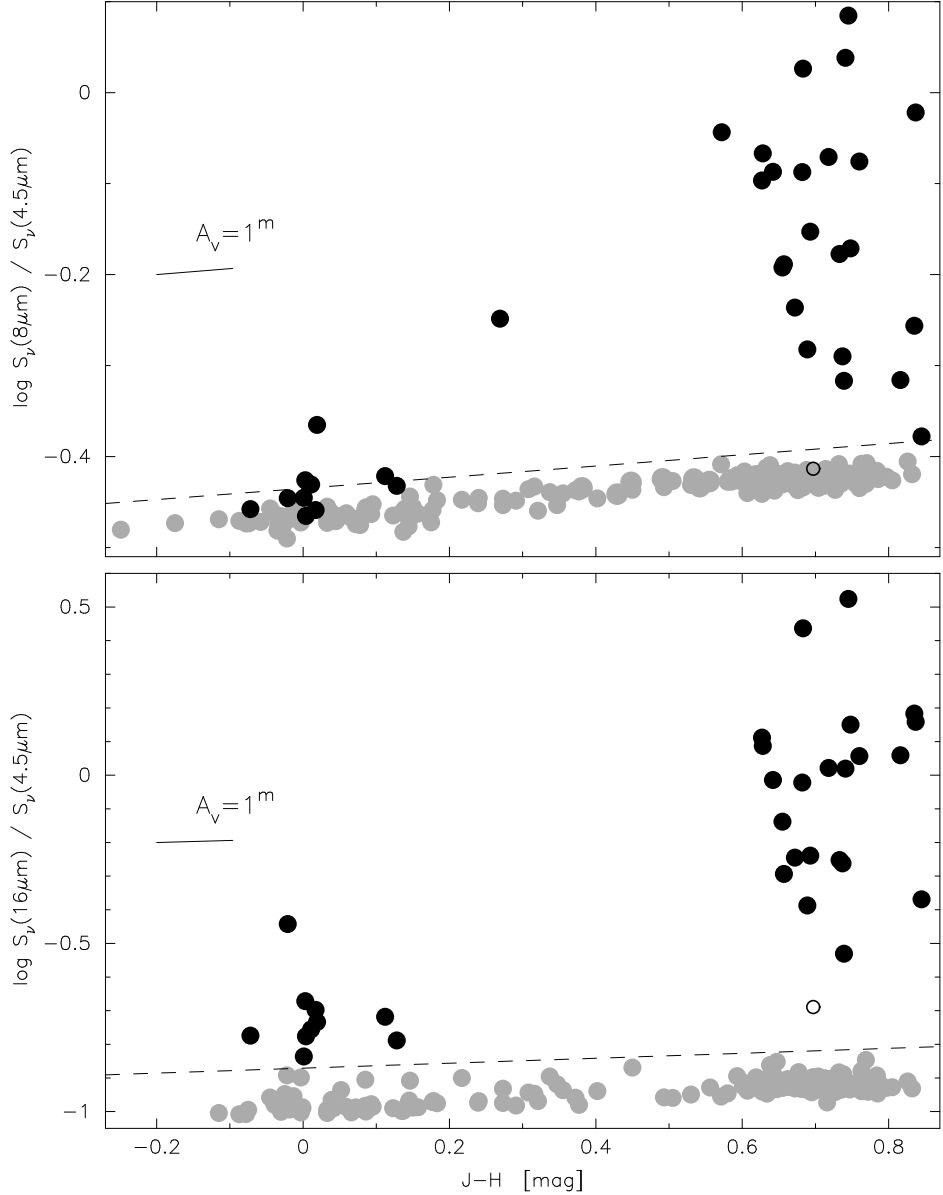}
\caption{
  \label{fig:ccd} Color-color diagrams showing $J-H$ along the abscissa, 
   tracing the stellar photosphere, and the 8\micron\ to 4.5\micron\ flux 
   ratio (top panel) and 16\micron\ to 4.5\micron\ flux ratio (bottom) along
   the ordinate, diagnostic of circumstellar disks. Dashed lines indicate the 
   thresholds adopted to identify sources with infrared excesses, corresponding
   to a color excess above the photosphere of 8\% (top panel) and 25\% (bottom).
   Black circles represent sources identified with a 8\micron\ or 16\micron\ 
   excess, and gray circles are sources without a detectable excess. 
   The open circle circle represents ScoPMS~17 for which the 16\micron\ 
   excess is questionable based on possible source confusion (see 
   Table~\ref{tbl:phot}).
   The internal uncertainties in the \Spitzer\ flux ratios are all 
   smaller than the symbol size, and the median $J-H$ uncertainty is 
   0.036~mag. The source $[$PBB2002$]$ USco J161420.2$-$190648, which has
   an excess at both 8\micron\ and 16\micron, is offscale on these plots. 
   The reddening vector from \citet{Mathis90} is indicated.
}
\end{figure}

The excess properties of the Upper Sco sources are not uniform across spectral 
type as demonstrated in Figure~\ref{fig:ccd}. The 8\micron\ excess fraction 
for K+M stars (24/127) is higher than for B+A stars (5/61) at 
the 92\% confidence level and higher than for F+G stars (0/30) at 99.2\% 
confidence as determined from the two-tailed Fisher's Exact Test. At 16\micron, 
the K+M excess fraction (23/121) is similar to that for B+A stars (10/52), but 
higher than for F+G stars (0/22) at 97.5\% confidence. 

More telling differences between early and late spectral types are observed 
in the magnitude of the excesses. Nine B+A stars have a $R_{16}$ color 
excess less than twice the stellar photosphere, while all K+M sources exceed 
this limit with excesses up to 27 times the photospheric level. Similarly, at 
8\micron, all but two B+A excess sources have a $R_8$ excess less than 10\% of 
the photosphere, while 23 of the 24 K+M excess sources have larger color 
excesses. 

\begin{figure}
\includegraphics[angle=0,scale=0.7]{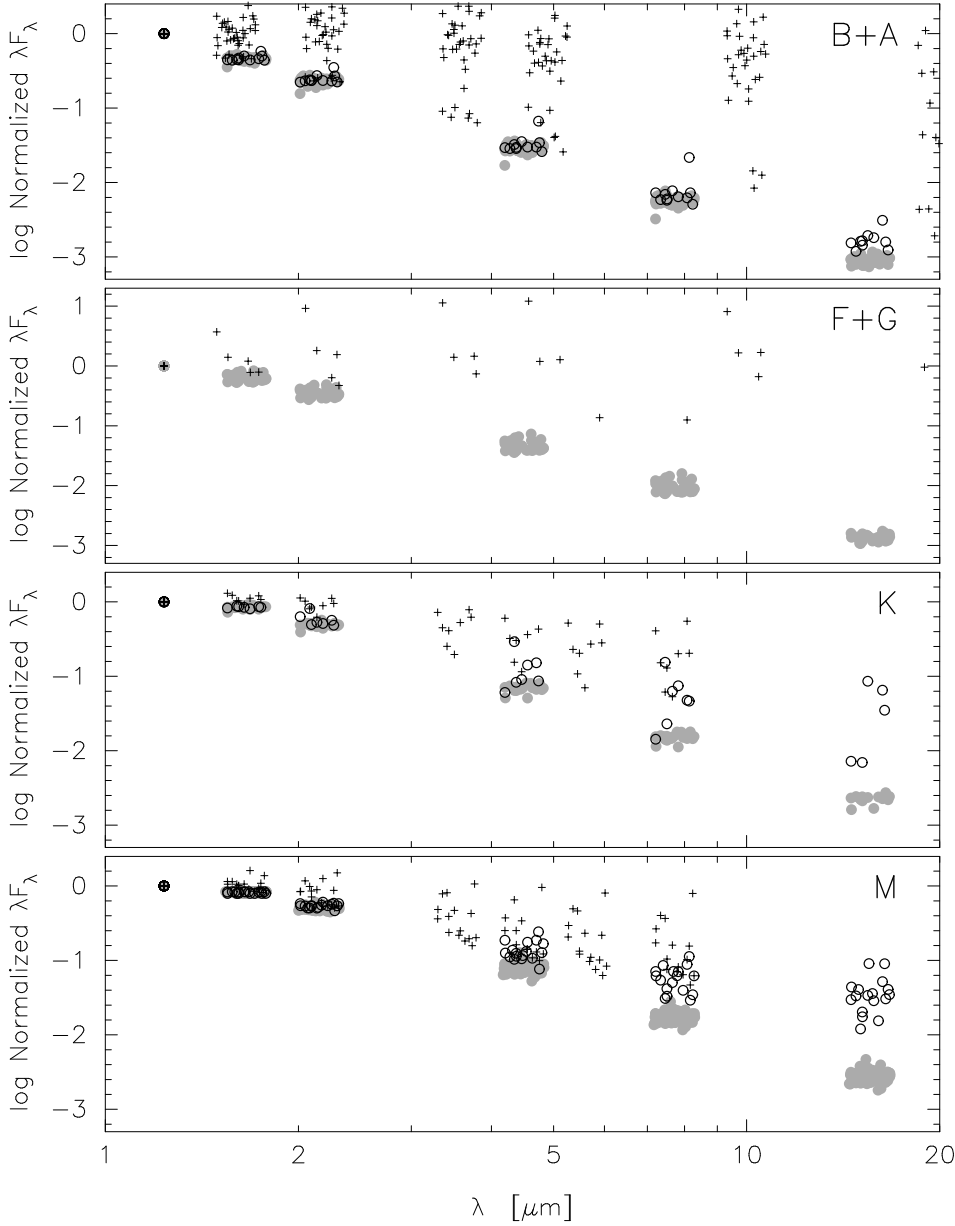}
\caption{
  \label{fig:sed} Dereddened spectral energy distributions for B+A (top), F+G,
  K, and M (bottom) stars. Gray circles represent Upper Sco sources 
  that do not have a detectable excess at wavelengths $\le$ 16\micron,
  and open circles are sources with an excess in one or more bands. 
  Plus-symbols represent Herbig Ae/Be stars (B, A, and F spectral types) and 
  Class~II sources in Taurus (G, K, and M spectral types) listed in 
  \citet{Hillenbrand92} and \citet{Hartmann05}. The SEDs have been normalized 
  to $J$-band, and a random offset has been added to the wavelengths to 
  illustrate the distribution of points.
}
\end{figure}

In Figure~\ref{fig:sed}, we assess the evolutionary state of the circumstellar 
disks in Upper Sco by comparing normalized, dereddened spectral energy 
distributions (SEDs) for stars in Upper Sco with a sample of well 
known T Tauri \citep{Hartmann05} and Herbig Ae/Be \citep{Hillenbrand92} stars 
that have tabulated photometry. The Taurus and Herbig Ae/Be objects are 
expected to represent young stars surrounded by primordial, optically thick, 
circumstellar accretion disks.

As shown in Figure~\ref{fig:sed}, the SEDs for B and A stars in Upper Sco
differ substantially from most Herbig Ae/Be stars. Herbig Ae/Be stars typically
have excesses at wavelengths as short as 2\micron\ and have fractional 
excess luminosities that are 10-100 times the photosphere at 
10-20\micron. By comparison, only one B or A star in Upper Sco has a $K$-band 
excess (HIP~78207), and the fractional excess luminosity at 16\micron\ is 
typically less than twice the photosphere. 

Surprisingly, none of the F and G stars in our Upper Sco sample exhibit a 
detectable excess. While \citet{Chen05} identified a 24\micron\ excess 
around one of five F+G stars in Upper Sco, and the G6\,V star HD~143006 is 
surrounded by an optically thick disk \citep{Sylvester96}, overall infrared 
excesses at wavelengths $\le$ 16\micron\ are relatively rare for this spectral 
type range at the age of Upper~Sco.

The results for the B, A, F, and G stars imply that the reservoir of small 
dust grains in a primordial, optically thick, inner disk have been largely
depleted by an age of 5~Myr for \about 1-20\msun\ stars. The inner-disk
radius inferred by the weak (or lack-of) excess emission at 16\micron\ is
\about 4-10~AU for 6000-10,000~K photospheres assuming optically thin, 
blackbody dust emission \citep{Jura03}.
These results are consistent with the low fraction of accreting B and A stars
found in the 4~Myr-old Trumpler~37 cluster \citep{Sicilia06} and the 5~Myr-old
Orion OBIb association \citep{Hernandez06}. However, \citet{Hernandez06} 
found that 7 of 11 F-stars in Orion~OBIb contain 24\micron\ excesses 
consistent with a debris disk, suggesting that the excess fraction for F 
and G-stars in Upper Sco may increase once our longer 
wavelength observations are obtained.

In contrast to the massive stars, the K and M stars in Upper Sco with 
infrared excesses have characteristics similar to optically thick primordial 
disks. As discussed above and demonstrated in Figure~\ref{fig:sed}, 
the K+M stars have larger fractional excesses than the B+A stars, and the
magnitude of the excesses overlaps with that observed toward Class~II stars in 
Taurus. A further connection between the disks in Upper Sco and Taurus is 
found by considering evidence for disk accretion as traced by H$\alpha$ 
emission. Of the 100 stars in our sample with measured H$\alpha$ line 
strengths (see references in \S\ref{sample}), nine have H$\alpha$ equivalent 
widths consistent with accretion according to the criteria recommended by 
\citet{White03}. Eight of these nine sources have an infrared excess 
and are therefore likely surrounded by accretion disks. We note however that 
the H$\alpha$ equivalent widths for 14 K+M stars with excesses are consistent 
with active chromospheres. These similarities suggest that many of the K+M 
stars in Upper Sco remain surrounded by optically thick disks. Assuming 
optically thick blackbody emission, the 4.5\micron\ and 8\micron\ excesses 
may imply the presence of dust at radii as small as \about 0.05~AU 
\citep{Jura03}. 

Differences in the excess characteristics between the \about 1-2 Myr-old 
Taurus and 5~Myr Upper Sco populations are notable since they may reflect
temporal evolution in the disk properties. While the magnitude of the excesses 
overlaps between the two samples, the excesses are larger on average for Taurus 
as seen in Figure~\ref{fig:sed}. Furthermore, about half of the stars in 
Taurus exhibit a $K$-band excess \citep{Strom89} compared to only two (1.5\%) 
K+M stars in Upper Sco ($[$PBB2002$]$~USco~J161420.2$-$190648 and 
$[$PZ99$]$~J160421.7$-$213028). Similarly, while 68\% of the stars in Taurus 
exhibit a 3.6\micron\ excess \citep{Haisch01}, only $19^{+5}_{-4}$\% of K+M 
stars in Upper Sco have a 8\micron\ excess. Therefore, not only are there 
fewer sources with disks in Upper Sco, but the disks that remain lack 
the hot dust found in younger stars. \citet{Sicilia06} found similar 
tendencies for the low mass population in Trumpler~37 compared to Taurus. They 
attribute these differences to grain growth or dust settling in the inner 
disk, although further observations and modeling are needed to explore 
the relevance of these ideas to stars in Upper Sco.

To summarize, 19\% of K0-M5 stars in Upper Sco possess infrared excesses 
similar to Class~II sources in Taurus, indicating primordial disks last around 
an appreciable number of 0.1-1\msun\ stars for at least 5~Myr. By contrast, 
only \aboutless 1\% of the more massive stars in Upper Sco contain such disks 
within an orbital radius of \about 4-10~AU. Similar results have been reported 
for the 4~Myr-old Trumpler~37 cluster \citep{Sicilia06}, and the 2-3~Myr-old 
IC~348 cluster may also contain a higher fraction of disks around
later type stars \citep{Lada06}. Our observations of Upper Sco extend these 
conclusions to the full range of stellar masses down to the hydrogen burning 
limit at an age of \about 5~Myr. These results establish that warm dust in
the terrestrial zone persist for longer times around stars with masses 
\aboutless 1\msun.

\acknowledgements

JMC would like to thank the Spitzer Science Center staff for 
patiently answering numerous questions regarding Spitzer data. EEM is 
supported by a Clay Postdoctoral Fellowship through the Smithsonian 
Astrophysical Observatory. MRM acknowledges support by NASA through the NASA
Astrobiology Institute under cooperative agreement CAN-02-OSS-02 issued 
through the Office of Space Science. This work is based on observations made 
with the {\it Spitzer Space Telescope}, which is operated by 
JPL/Caltech under a contract with NASA. Support for this work was provided by 
NASA through an award issued by JPL/Caltech. 
This research made use of the 
SIMBAD database, operated at CDS, Strasbourg, France. This paper made use of 
data products from the Two Micron All Sky Survey, which is a joint project of 
the U. Massachusetts and the Infrared Processing and Analysis Center/Caltech, 
funded by NASA and the NSF.

\rotate

\begin{deluxetable}{lrrcc@{\extracolsep{10pt}}cccccc}
\tabletypesize{\scriptsize}
\tablewidth{525pt}
\tablecaption{Observed IRAC and IRS Fluxes\label{tbl:phot}}
\tablehead{
  \colhead{Source}  &
  \colhead{SpT}     &
  \colhead{SpT Ref}     &
  \multicolumn{2}{c}{IRAC 4.5\micron\tablenotemark{a}}  &
  \multicolumn{2}{c}{IRAC 8\micron\tablenotemark{b}}  &
  \multicolumn{2}{c}{IRS 16\micron\tablenotemark{c,d}} &
  \multicolumn{2}{c}{Excess?}\\
  \cline{4-5}
  \cline{6-7}
  \cline{8-9}
  \cline{10-11}
  \colhead{}    &
  \colhead{}    &
  \colhead{}    &
  $S_\nu$ (mJy) &
  $\sigma$ (mJy) &
  $S_\nu$ (mJy) &
  $\sigma$ (mJy) &
  $S_\nu$ (mJy) &
  $\sigma$ (mJy) &
  8\micron &
  16\micron
}
\startdata
\multicolumn{11}{c}{{\it a) Upper Sco sources from this study}}\\
HD 142987                 & G4         & MML    &  \dataset[ADS/Sa.Spitzer#13867264]{181.3}  &      2.2  &  \dataset[ADS/Sa.Spitzer#13867264]{67.9}  &      0.4  &  \dataset[ADS/Sa.Spitzer#13888000]{19.9}  &      0.1 & N & N\\
HD 147810                 & G1         & MML    &  \dataset[ADS/Sa.Spitzer#13864704]{228.9}  &      2.8  &  \dataset[ADS/Sa.Spitzer#13864704]{82.0}  &      0.8  &  \dataset[ADS/Sa.Spitzer#13885952]{26.3}  &      0.2 & N & N\\
HD 149598                 & G0         & MML    &  \dataset[ADS/Sa.Spitzer#13870336]{113.0}  &      1.4  &  \dataset[ADS/Sa.Spitzer#13870336]{41.4}  &      0.3  &  \dataset[ADS/Sa.Spitzer#13889280]{12.9}  &      0.1 & N & N\\
HIP 76071                 & B9V        & H88    &  \dataset[ADS/Sa.Spitzer#13861888]{277.0}  &      3.4  &  \dataset[ADS/Sa.Spitzer#13861888]{93.2}  &      0.8  &  \dataset[ADS/Sa.Spitzer#13882624]{27.4}  &      0.2 & N & N\\
HIP 76310                 & A0V        & H88    &  \dataset[ADS/Sa.Spitzer#13863424]{247.6}  &      3.0  &  \dataset[ADS/Sa.Spitzer#13863424]{88.8}  &      0.8  &  \dataset[ADS/Sa.Spitzer#13884416]{89.4}  &      0.5 & N & Y\\
HIP 76633                 & B9V        & H88    &  \dataset[ADS/Sa.Spitzer#13866240]{179.4}  &      2.2  &  \dataset[ADS/Sa.Spitzer#13866240]{63.2}  &      0.4  &  \dataset[ADS/Sa.Spitzer#13887488]{19.9}  &      0.1 & N & N\\
HIP 77457                 & A7IV       & H82    &  \dataset[ADS/Sa.Spitzer#13864960]{224.4}  &      2.7  &  \dataset[ADS/Sa.Spitzer#13864960]{77.0}  &      0.9  &  \dataset[ADS/Sa.Spitzer#13886208]{23.0}  &      0.1 & N & N\\
\enddata
\tablenotetext{a}{IRAC 4.5\micron\ photometry measured using a flux calibration factor of 0.1388 MJy/sr per DN/s.}
\tablenotetext{b}{IRAC 8\micron\ photometry measured using a flux calibration factor of 0.2021 MJy/sr per DN/s.}
\tablenotetext{c}{IRS 16\micron\ PUI photometry measured using a flux calibration factor of 0.01375 MJy/sr per $e^{-}$/s.}
\tablenotetext{d}{Saturated sources are listed as ``sat"}
\tablenotetext{e}{Aperture curve of growth deviates from a point source by 
   more than 4\%.}
\tablenotetext{f}{ScoPMS~17 is not considered an excess source for this study
   since the apparent 16\micron\ excess is likely from a nearby source that 
   falls partially within the mesurement aperture.}
\tablerefs{H82: \citet{Houk82}; H88: \citet{Houk88}; K99: \citet{Kunkel99}; K00: \citet{Kohler00}; MML: Mamajek, Meyer, \& Liebert, unpublished; P98: \citet{Preibisch98}; P01: \citet{Preibisch01}; P02: \citet{Preibisch02}; W94: \citet{Walter94}}
\end{deluxetable}


\end{document}